\documentclass[aps,prd,floatfix,nofootinbib,twocolumn]{revtex4}
\usepackage{graphicx}
\usepackage{amsmath}
\usepackage{amssymb}
\usepackage{subfigure}
\usepackage{bm}
\newcommand{\beq}{\begin{equation}}
\newcommand{\eeq}{\end{equation}}
\newcommand{\bea}{\begin{eqnarray}}
\newcommand{\eea}{\end{eqnarray}}
\newcommand{\MeV}{\;\text{MeV}}

\begin{document}
\title{ Dual condensates at finite isospin chemical potential }

\author{Zhao~Zhang}\email{zhaozhang@pku.org.cn}
\author{Qing~Miao}
\affiliation{ School of Mathematics and Physics, North China
Electric Power University, Beijing 102206, China}

\begin{abstract}

The dual observables as order parameters for center symmetry are tested at finite isospin chemical potential $\mu_I$
in a Polyakov-loop enhanced chiral model of QCD with physical quark masses. As a counterpart of the dressed Polyakov-loop, the
first Fourier moment of pion condensate is introduced for $\mu_I>{m_\pi}/{2}$ under the temporal twisted boundary
conditions for quarks. We demonstrate that this dual condensate exhibits the similar temperature dependence as the
conventional Polyakov-loop. We confirm that its rapid increase with $T$ is driven by the evaporating of
pion condensation. On the other hand, the dressed Polyakov-loop shows abnormal thermal behavior, which even
decreases with $T$ at low temperatures due to the influence of pion condensate. We thus argue that in QCD
the critical temperature extracting from a dual observable may have nothing to do with the quark
confinement-deconfinement transition if the quark mass is very small.

\vspace{10pt} PACS number(s): 12.38.Aw; 11.30.RD; 12.38.Lg;
\end{abstract}

\date{\today}
\maketitle

\section{Introduction}

The main phenomena in QCD at finite temperature and density are the chiral restoration and deconfining
phase transitions. In the chiral limit, the standard order parameter for chiral transition is the quark
condensate. However, it is conceptually difficult to define an order parameter for deconfinement in
QCD. Usually, the expectation value of the Polyakov loop (PL) is adopted as the indicator for quark
deconfining transition. This quantity is a true order parameter for center symmetry in pure Yang-Mills theory.
Nevertheless, this symmetry is badly broken by the light dynamical quarks in QCD. These two order parameters 
had been extensively studied in lattice QCD. It suggests that both phase transitions are smooth crossovers 
at finite $T$ and zero density and two pseudo critical temperatures are very close to each
other \cite{Aoki:2009sc,Borsanyi:2010bp}.

It is well known that the nonzero Dirac zero-mode density is responsible for the dynamical chiral symmetry
breaking in QCD, according to the celebrated Banks-Casher relation \cite{Banks:1980}. An interesting question
is to what extent the spectral properties of the Dirac operator contain the confinement information. Recently,
some authors have tried to link the Dirac spectral modes to the PL or its equivalent quantities with
the same winding number in the time direction
\cite{Gattringer:2006ci,Bilgici:2008qy,Bilgici:2009tx,Zhang:2010ui,Synatschke:2007yt,Synatschke:2008yt,Braun:2009gm}.
In these studies, some dual observables are introduced as the new order parameters for center symmetry by
using the twisted boundary conditions for quarks. Especially, it is demonstrated in the formalism of lattice
QCD \cite{Bilgici:2008qy,Bilgici:2009tx,Zhang:2010ui} that the dressed polyakov-loop (DPL) interpolates between the
chiral condensate and the thin PL. The studies from the functional methods
\cite{Fischer:2009gk,Fischer:2009wc,Fischer:2010fx,Fischer:2011mz} and effective models
\cite{Kashiwa:2009ki,Gatto:2010qs,Mukherjee:2010cp} also suggest the DPL shows the order parameter-like
behavior like the thin PL.

In principle, one can construct many dual observables which transform in the same way as the thin PL under
the center transformation \cite{Braun:2009gm,Zhang:2010ui}. They are true order parameters for deconfinement in the static
limit $m\rightarrow\infty$, where $m$ is the mass of dynamical quarks. However, the center symmetry is seriously
broken in QCD since the light quark masses are very small. Then, a question naturally arises: to what extent do these
quantities still contain the confinement information? Recently, it is demonstrated in the NJL model that the
rapid rise of DPL near $T_c$ is totally driven by the chiral transition \cite{Benic:2013zaa}. The author argues that 
the reason for it is the lacking confinement of NJL. Note that such an explanation is not so convincing since the thermal behavior 
of DPL calculated in NJL is quite similar to that obtained in other methods. So even the argument is only against the NJL 
result, the main conclusion that the DPL does not reflect the deconfinement transition in \cite{Benic:2013zaa} 
may also be true in QCD.

To further test weather the dual observables can be used as order parameters, we extend the previous study to 
finite isospin chemical potential $\mu_I$ by simultaneously considering the pion condensation and the twisted 
boundary conditions in this paper.
We mainly concentrate on the thermal properties of two simple dual observables for $\mu_I>m_\pi/2$
\footnote{It is well known that the charged pion condensation appears at low temperatures for
$\mu_I>m_\pi/2$ in QCD \cite{Son:2001}.}: the DPL and the first Fourier moment of the generalized pion condensate.
Here we refer to the later as the dual pion condensate (DPC), which is the counterpart of the DPL at finite
$\mu_I$. Due to the influence of pion condensate, the thermal property of DPL may change explicitly for
$\mu_I>m_\pi/2$. Thus it is interesting to check whether the DPL still behaves like an order parameter in this situation.
Second, similar to the DPL, the DPC transforms in the same manner as the thin PL under the center transformation.
So it is also interesting to explore whether this simple dual quantity can be used to indicate the
deconfinement transition at finite isospin density.

We employ the PL enhanced NJL model (PNJL) in our investigation by adopting the U(1)-valued boundary conditions.
Compared to \cite{Mukherjee:2010cp,Benic:2013zaa}, the advantage of PNJL is that the PL dynamics is included to 
partially mimic the confinement. Moreover, the pion condensate and PL obtained in lattice simulations \cite{Kogut:2002,Kogut:2004}
for $\mu_I>m_\pi/2$ can be well reproduced in this model \cite{Zhang:2006gu}. The paper is organized as follows.
In Sec.II, the dual pion condensate is defined and the PNJL model with the twisted boundary conditions for
$\mu_I>m_\pi/2$ is introduced. The numerical results and discussion are given in Sec.III. In Sec.IV, we summarize.


\section{ Dual pion condensate and PNJL model with twisted boundary condition for $\mu_I>m_\pi/2$ }

\subsection{ Dual pion condensates for $\mu_I>m_\pi$/2 }

According to \cite{Bilgici:2008qy}, the dual quark condensates are defined as
\beq
\Sigma^{(n)}_{\sigma}=-\int^{2\pi}_0\frac{d\phi}{2\pi}e^{-in\phi}\sigma(\phi)\label{dualquarkc},
\eeq
where $n$ is an integer and $\sigma(\phi)$ is the generalized quark condensate
\beq
\sigma(\phi)=\langle{\bar{\psi}\psi\rangle}_{\phi}=-\frac{1}{V}\langle{Tr[(m+D_{\phi})^{-1}]\rangle}\label{quarkc},
\eeq
which is obtained with the twisted boundary condition in the time direction
\beq
\psi(x,\beta=1/T)=e^{i\phi}\psi(x,0)\label{twistedbc}.
\eeq
The $D_{\phi}$ in \eqref{quarkc} is the Dirac operator without the quark mass for the twisted angle $\phi$. Note
that $\phi=\pi$ corresponds to the physical boundary condition. The dressed PL is defined as the first Fourier
moment of $\sigma(\phi)$, namely
\beq
\Sigma^{(1)}_{\sigma}=-\int^{2\pi}_0\frac{d\phi}{2\pi}e^{-i\phi}\sigma(\phi).
\eeq
In the lattice language, this quantity only includes the contributions of (infinite) closed loops with the winding
number one in the compact time direction \cite{Bilgici:2008qy}. So it belongs to the same class as the thin PL under
the center transformation.

In the same way, we can introduce the dual pion condensates at finite $\mu_I$.
The charged pion condensates for $\mu_I>m_\pi/2$ are defined as
\begin{equation}
\langle{\bar{\psi}i\gamma_5\tau_{+}\psi}\rangle=\pi^{+}=\frac{\pi}{\sqrt{2}}e^{i\theta},
\quad
\langle{\bar{\psi}i\gamma_5\tau_{-}\psi}\rangle=\pi^{-}=\frac{\pi}{\sqrt{2}}e^{-i\theta},
\label{pionc}
\end{equation}
where $\tau_{\pm}=(\tau_1\pm\tau_2)/\sqrt{2}$ and $\tau_i$ is the Pauli matrix in quark flavor space.
In \eqref{pionc}, nonzero $\pi$ indicates the spontaneous breaking of the isospin $I_3$ symmetry and
the breaking direction is described by the phase factor $\theta$. Without loss of generality, we adopt
$\theta=0$ in the following and the pion condensate is expressed as
\begin{equation}
\langle{\bar{\psi}i\gamma_5\tau_1\psi}\rangle=\pi.
\end{equation}
Similar to \eqref{dualquarkc}, we can define the dual pion condensates
\beq
\Sigma^{(n)}_{\pi}=-\int^{2\pi}_0\frac{d\phi}{2\pi}e^{-in\phi}\pi(\phi),
\eeq
where $\pi(\phi)$ is the generalized pion condensate under the boundary condition
\eqref{twistedbc}, which takes the form
\beq
\pi(\phi)=-\frac{1}{V}\langle{Tr[i\gamma_5\tau_1(m+D_{\phi})^{-1}]\rangle} \label{piongen}.
\eeq
As mentioned, the DPC is defined as the first Fourier moment of $\pi(\phi)$, namely
\beq
\Sigma^{(1)}_{\pi}=-\int^{2\pi}_0\frac{d\phi}{2\pi}e^{-i\phi}\pi(\phi).
\eeq

Analogous to the DPL (and also the dual density proposed in \cite{Braun:2009gm}), $\Sigma^{(1)}_{\pi}$
is gauge invariant which merely includes the contributions of closed loops with wingding number one.
Thus it is another simple dual observable transforming in the same manner as the thin PL under
the Z(3) center transformation. It is interesting to check whether this quantity also exhibits an
order parameter-like behavior with increasing T at finite isospin density.

The previous studies \cite{Kogut:2004,Zhang:2006gu} suggest that under the physical boundary condition,
the pion condensate competes with the quark condensate for $\mu_I>m_\pi/2$. Or in other words, the quark
condensate partially rotates into the pion condensate when the isospin chemical potential surpasses the half
of the pion mass at zero $T$ and their competition becomes more involved at finite $T$. We can
expect that there may exist the similar interplay between these two condensates for
other twisted boundary angles. This implies $\pi(\phi)$ affects $\sigma(\phi)$, and vice versa.
So the thermal behavior of DPL at $\mu_I>m_\pi/2$ may deviate significantly from that at zero $\mu_I$ due to the
influence of $\pi(\phi)$. We will test whether such a deviation still supports the DPL as an indicator
for quark deconfinement transition with physical quark masses.

\subsection{ PNJL model for $\mu_I>m_\pi$ with twisted boundary condition }

We adopt the following lagrangian of two-flavor PNJL model

\begin{eqnarray}
\mathcal{L}&=&\bar{\psi}\left(i\gamma_{\mu}D^{\mu}+\gamma_0\hat{\mu}
-\hat{m}_0-i{\lambda}\gamma_5\tau_1\right)\psi\nonumber\\
&&+g_s\left[\left(\bar{\psi}\psi\right)^2+\left(\bar{\psi}i\gamma_5\vec{\tau}\psi \right)^2\right]
-g_v^s\left(\bar{\psi}\gamma_\mu\psi\right)^2\nonumber\\
&&-g_v^v\left(\bar{\psi}\vec{\tau}\gamma_\mu\psi\right)^2
-\mathcal{U}\left(\Phi,\bar{\Phi},T\right), \label{lagr}
\end{eqnarray}
where the last term is the effective PL potential. This type model has been used to study the DPL
at zero density \cite{Kashiwa:2009ki}. Compared to \cite{Kashiwa:2009ki}, we ignore the eight-quark interaction
but include four-quark vector interactions with two different couplings. It is demonstrated in \cite{Zhang:2013oia}
that the mismatch between $g_v^v$ and $g_v^s$ can lead to non-anomaly flavor mixing at finite baryon and isospin
densities.

The $\hat{m_0}$ is the matrix of current quark masses
\begin{equation}
\hat{m_0}=\bigg(\begin{array}{cc}
    m_u & \\
     & m_d\end{array}
     \bigg),
\end{equation}
and we choose $m_u = m_d \equiv m$.
The $\hat{\mu}$ is the matrix of quark chemical potentials
\begin{equation}
\hat{\mu}=\bigg(\begin{array}{cc}
    \mu_u & \\
     & \mu_d\end{array}
 \bigg)=\bigg(\begin{array}{cc}
    \mu+\mu_I & \\
     & \mu-\mu_I\end{array}
 \bigg),
\end{equation}
with
\begin{eqnarray}
\mu=\frac{\mu_u+\mu_d}{2}=\frac{\mu_B}{3}~~~~~~\mathrm{and}~~~~~~\mu_I=\frac{\mu_u-\mu_d}{2} \label{chemicalpts}.
\end{eqnarray}
The $\mu_B$ and $\mu_I$ in \eqref{chemicalpts} are the baryon and isospin chemical potentials, which
correspond to the conserved baryon and isospin charges, respectively. Following \cite{Kogut:2004,Zhang:2006gu}, we introduce
a small parameter $\lambda$ in \eqref{lagr}, which explicitly breaks the $I_3$ symmetry.

The mean field thermal potential of PNJL model for $\mu_I>m_\pi/2$ under the physical boundary condition has been given
in \cite{Zhang:2006gu}, where vector interactions are ignored. Its form is slightly modified when considering the
vector interactions
\begin{eqnarray}\label{thermalp}
&&\Omega={\cal{U}}(\Phi,\bar{\Phi},T)-2N_c\int{\frac{d^3p}{(2\pi)^3}}\big[E_p^-+E_p^+\big]\theta(\Lambda^2-\vec{p}^2)\nonumber\\
&&-2T\int{\frac{d^3p}{(2\pi)^3}}\Big\{\ln\Big[1+3\big(\Phi+\bar{\Phi}\mathrm{e}^{-\left(E_p^--\mu'\right)\beta}\big)\mathrm{e}^{-\left(E_p^--\mu'\right)\beta}
\nonumber\\
&&+\mathrm{e}^{-3\left(E_p^--\mu'\right)\beta}\Big]+\ln\Big[1+3\big(\bar{\Phi}+\Phi\mathrm{e}^{-\left(E_p^-+\mu'\right)\beta}\big)\mathrm{e}^{-\left(E_p^-+\mu'\right)\beta}
\nonumber\\
&&+
\mathrm{e}^{-3\left(E_p^-+\mu'\right))\beta}\Big]+\ln\Big[1+3\big(\Phi+\bar{\Phi}\mathrm{e}^{-\left(E_p^+-\mu'\right)\beta}\big)\mathrm{e}^{-\left(E_p^+-\mu'\right)\beta}
\nonumber\\
&&+
\mathrm{e}^{-3\left(E_p^+-\mu'\right)\beta}\Big]+\ln\Big[1+3\big(\bar{\Phi}+\Phi\mathrm{e}^{-\left(E_p^++\mu'\right)\beta}\big)\mathrm{e}^{-\left(E_p^++\mu'\right)\beta}
\nonumber\\
&&+
\mathrm{e}^{-3\left(E_p^++\mu'\right)\beta}\Big]\Big\}+g_s(\sigma^2+\pi^2)-g_v^s(\rho_u+\rho_d)^2
\nonumber\\
&&
-g_v^v(\rho_u-\rho_d)^2,\label{thermp}
\end{eqnarray}
with the quasi particle energy $E_p^{\pm}=\sqrt{(E_p\pm\mu_I')^2+N^2}$ and $E_p=\sqrt{\vec{p}^2+M^2}$
in which the two energy gaps are defined as
\begin{eqnarray}
M=m-2g_s\sigma,\\
N=\lambda-2g_s\pi.
\end{eqnarray}
The $\mu'$ and $\mu_I'$ are the shifted quark and isospin chemical potentials
\begin{eqnarray}
\mu'=\mu-2g_v^s(\rho_u+\rho_d),\quad
\mu_I'=\mu_I-2g_v^v(\rho_u-\rho_d),\label{shiftedcp}
\end{eqnarray}
where $\rho_{u(d)}$ is the u(d) quark density.

In the following, we only consider the situation with finite $\mu_I$  and zero $\mu$. In this case, the baryon number density is
zero for $\phi=\pi$ and $\Phi$ equals to $\bar{\Phi}$ strictly. Minimizing the thermal dynamical potential (\ref{thermalp}),
the motion equations for the mean fields $\sigma$, $\pi$, $\Phi$ and the density $\rho_I$ are determined through the
coupled equations
\begin{equation}
\frac{\partial\Omega}{\partial\sigma}=0,\quad\frac{\partial\Omega}{\partial\pi}=0,
\quad\frac{\partial\Omega}{\partial\Phi}=0,\quad\frac{\partial\Omega}{\partial\rho_I}=0.\label{coupleeqs1}
\end{equation}
This set of equations is then solved for the fields $\sigma$, $\pi$, $\Phi$ and $\rho_I$ as functions of $T$ and $\mu_I$.

Under the generalized boundary condition, the modified quark chemical potential $\mu'$ in
\eqref{thermalp} should be replaced by $iT(\phi-\pi)$ \cite{Bilgici:2008qy,Braun:2009gm,Kashiwa:2009ki},
which is nothing but an effective imaginary chemical potential. Strictly speaking, the $\mu'$
for $\phi\neq\pi$ should also contain the density-related contribution $2g_v^s(\rho_u+\rho_d)$ even the real
$\mu$ is zero. This is because the imaginary chemical potential also leads to a nonzero baryon
number density. It has been shown in \cite{Kashiwa:2009ki} that the coupling $g_v^s$ only has significant effect
on $\Sigma^{(1)}_{\sigma}$ for $T>1.5T_c$ in PNJL. Since we are only interested in the thermal
behavior of dual observables near and below $T_c$, the correction $2g_v^s(\rho_u+\rho_d)$ is ignored in our
calculation. Note that the $\mu_I'$ is still real and keeps the form as \eqref{shiftedcp}. The reason is
that it is the difference between $\mu_u'$ and $\mu_d'$ and their imaginary parts cancel each other
out for $\phi\neq\pi$.

According to the definition of DPL(and also the DPC), the twisted boundary condition is imposed on the Dirac
operator $D_{\phi}$, and the bracket $\langle\cdot\cdot\cdot\rangle$ still keeps the antiperiodic condition
with $\phi=\pi$ \cite{Bilgici:2008qy,Bilgici:2009tx}. So in our calculation, the $\Phi$ as a function of
$T$ and $\mu_I$ is first obtained by solving \eqref{coupleeqs1} using the physical boundary condition.
The other quantities, such as $\sigma(\phi)$, $\pi(\phi)$ and $\rho(\phi)$ are then determined by the
following coupled equations
\begin{equation}
\frac{\partial\Omega}{\partial\sigma(\phi)}=0,\quad\frac{\partial\Omega}{\partial\pi(\phi)}=0,
\quad\frac{\partial\Omega}{\partial\rho_I(\phi)}=0,\label{coupleeqs2}
\end{equation}
with $\Phi$ keeping its value for $\phi=\pi$. Such a treatment is consistent with
\cite{Kashiwa:2009ki}.

\begin{figure}[t]
\hspace{-.0\textwidth}
\begin{minipage}[t]{.45\textwidth}
\includegraphics*[width=\textwidth]{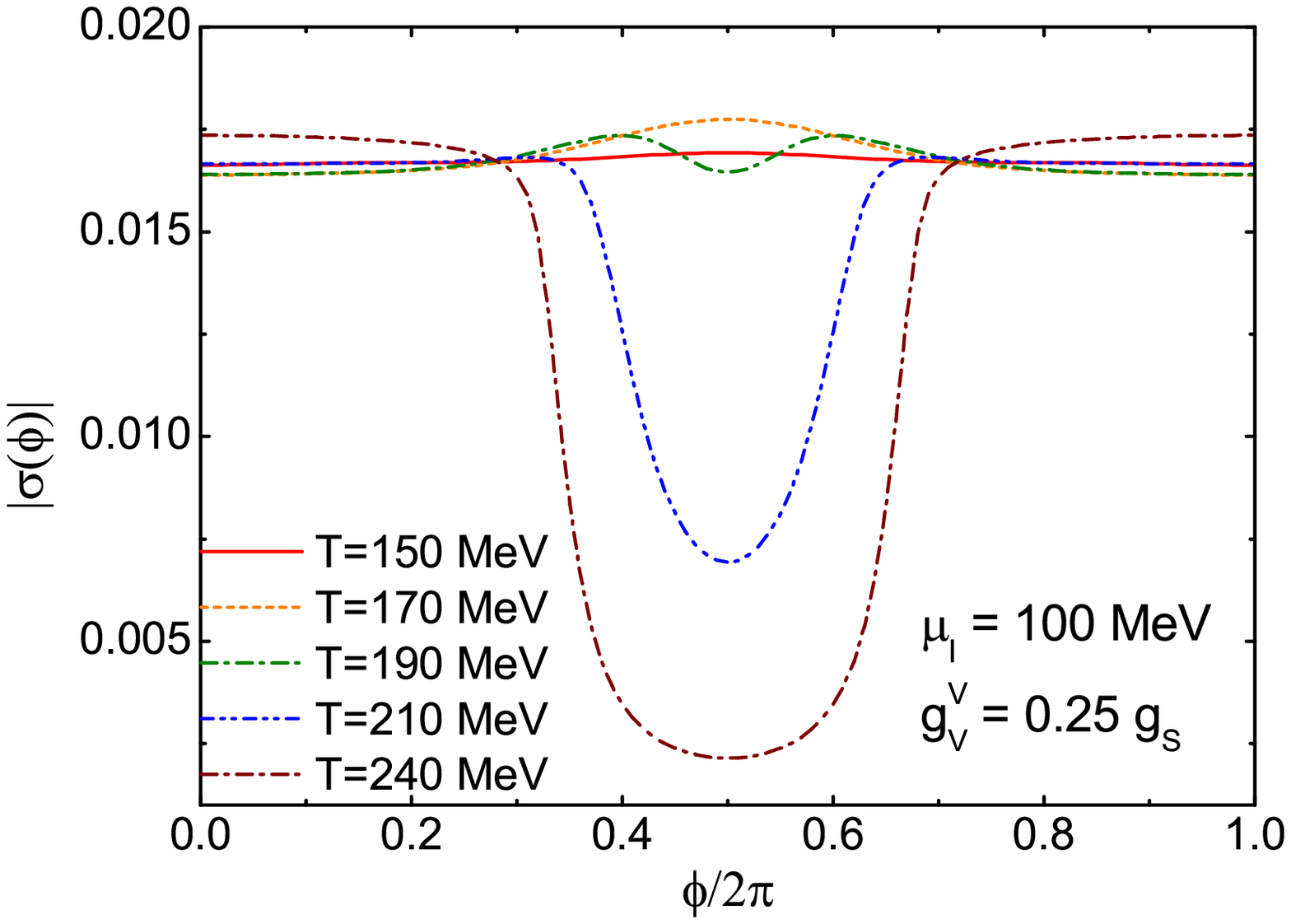}
\centerline{(a) }
\end{minipage}
\hspace{-.05\textwidth}
\begin{minipage}[t]{.45\textwidth}
\includegraphics*[width=\textwidth]{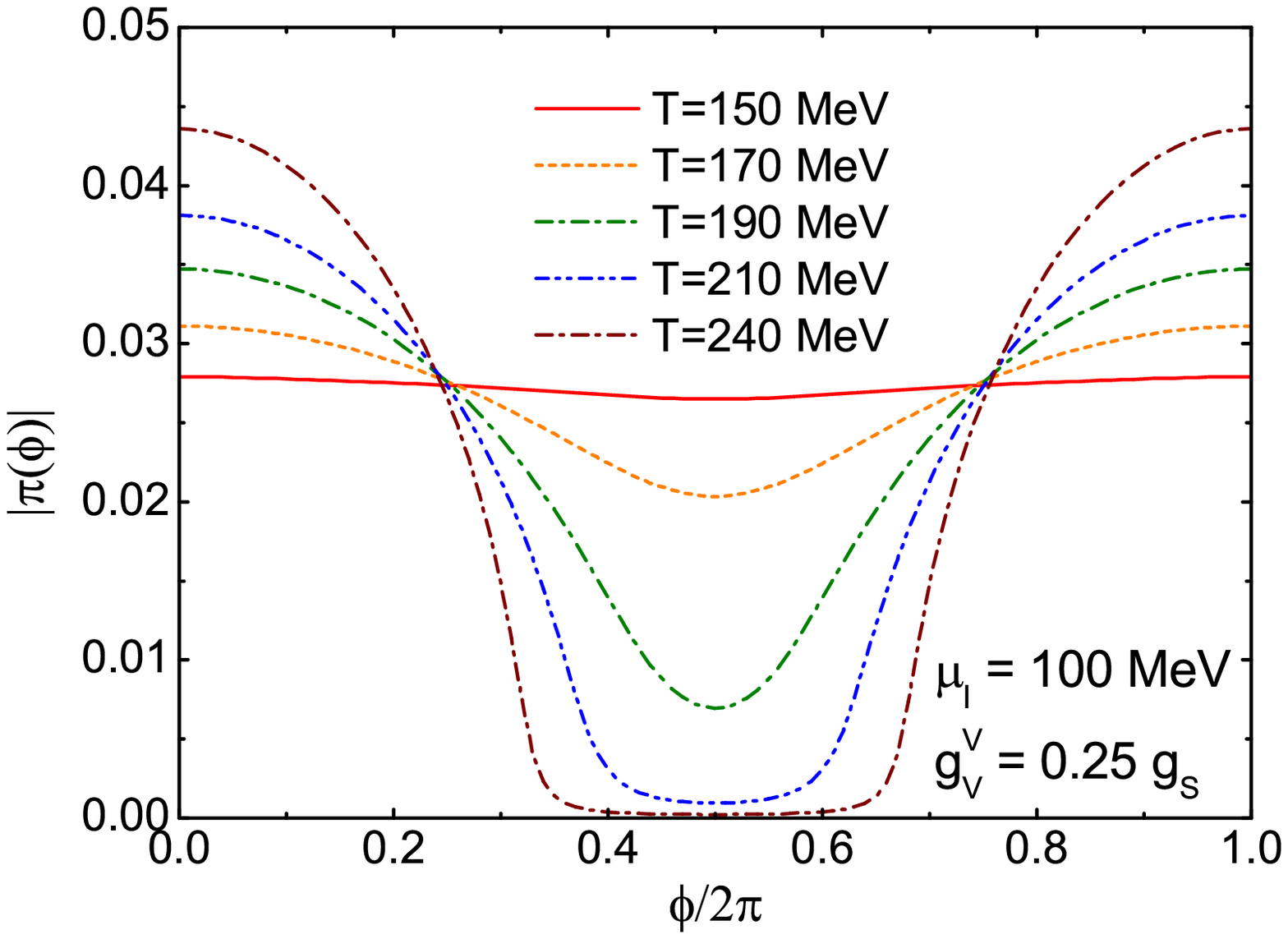}
\centerline{(b) }
\end{minipage}
\caption{ The twisted angle dependences of the quark condensate $\sigma({\phi})$ and pion condensate $\pi({\phi})$
at $\mu_I=100 \MeV$ for different temperatures. }
\label{fig:dualc1}
\end{figure}

\begin{figure}[t]
\hspace{-.0\textwidth}
\begin{minipage}[t]{.45\textwidth}
\includegraphics*[width=\textwidth]{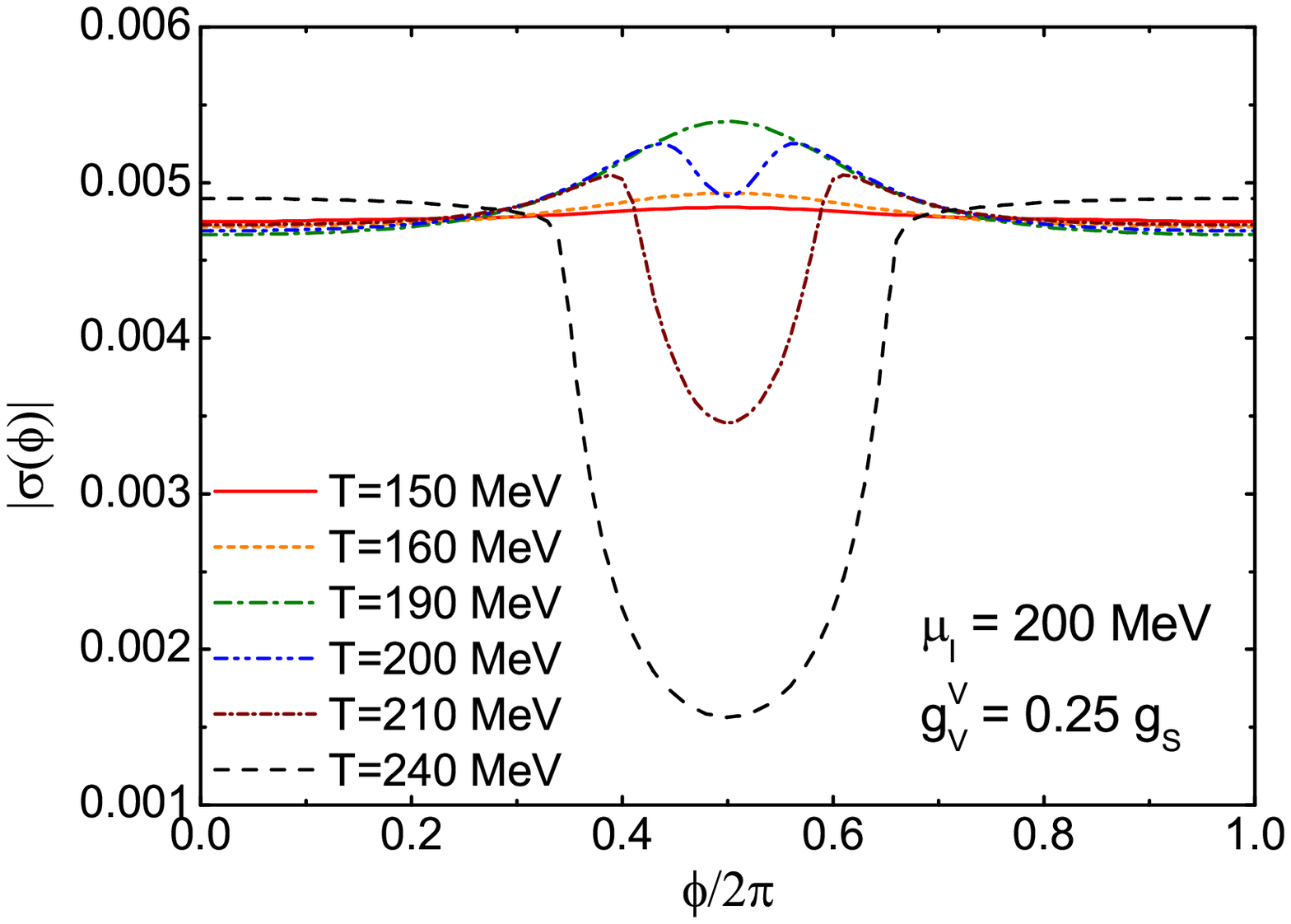}
\centerline{(a) }
\end{minipage}
\hspace{-.05\textwidth}
\begin{minipage}[t]{.45\textwidth}
\includegraphics*[width=\textwidth]{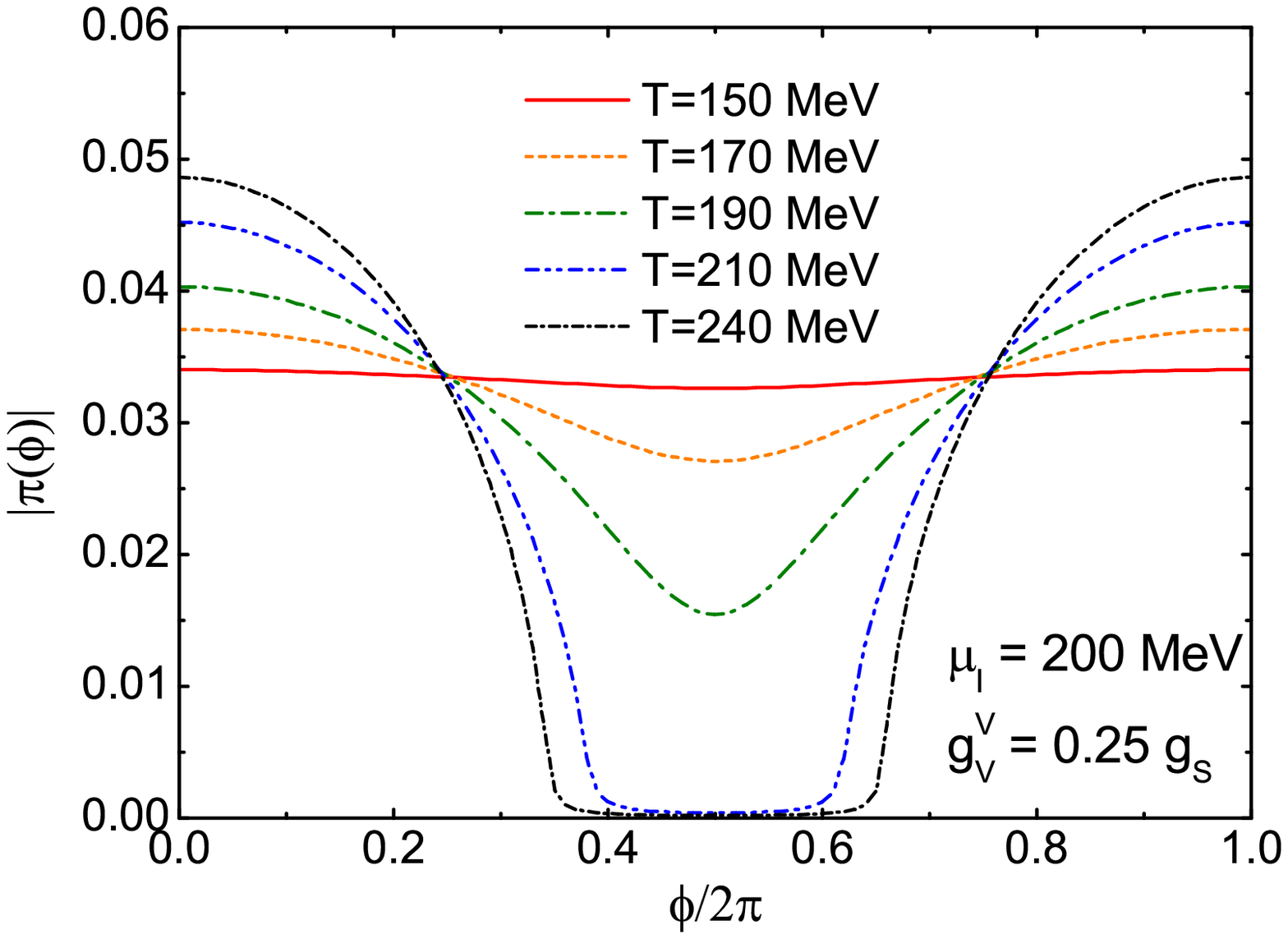}
\centerline{(b) }
\end{minipage}
\caption{ The twisted angle dependences of the quark condensate $\sigma({\phi})$ and pion condensate $\pi({\phi})$
at $\mu_I=200 \MeV$ for different temperatures. }
\label{fig:dualc2}
\end{figure}

\begin{figure}[t]
\begin{center}
\includegraphics[width=1.0\columnwidth]{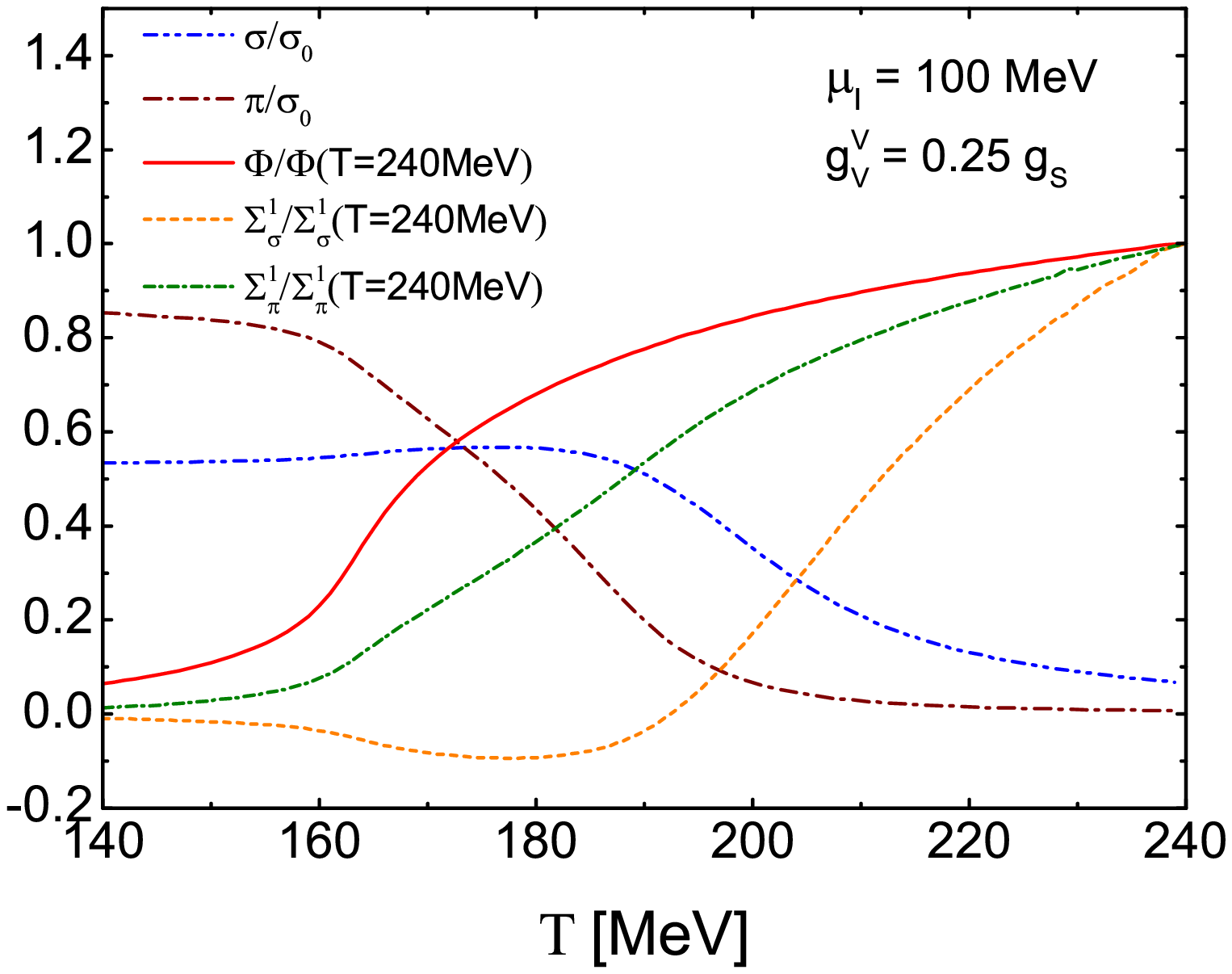} \vspace{1em}\\
\end{center}
\caption{ The temperature dependences of the normalized conventional ployakov loop,
quark and pion condensates and their corresponding dual parters at $\mu_I=100 \MeV$.
 }
\label{fig:plcondensate1}
\end{figure}

\begin{figure}[t]
\begin{center}
\includegraphics[width=1.0\columnwidth]{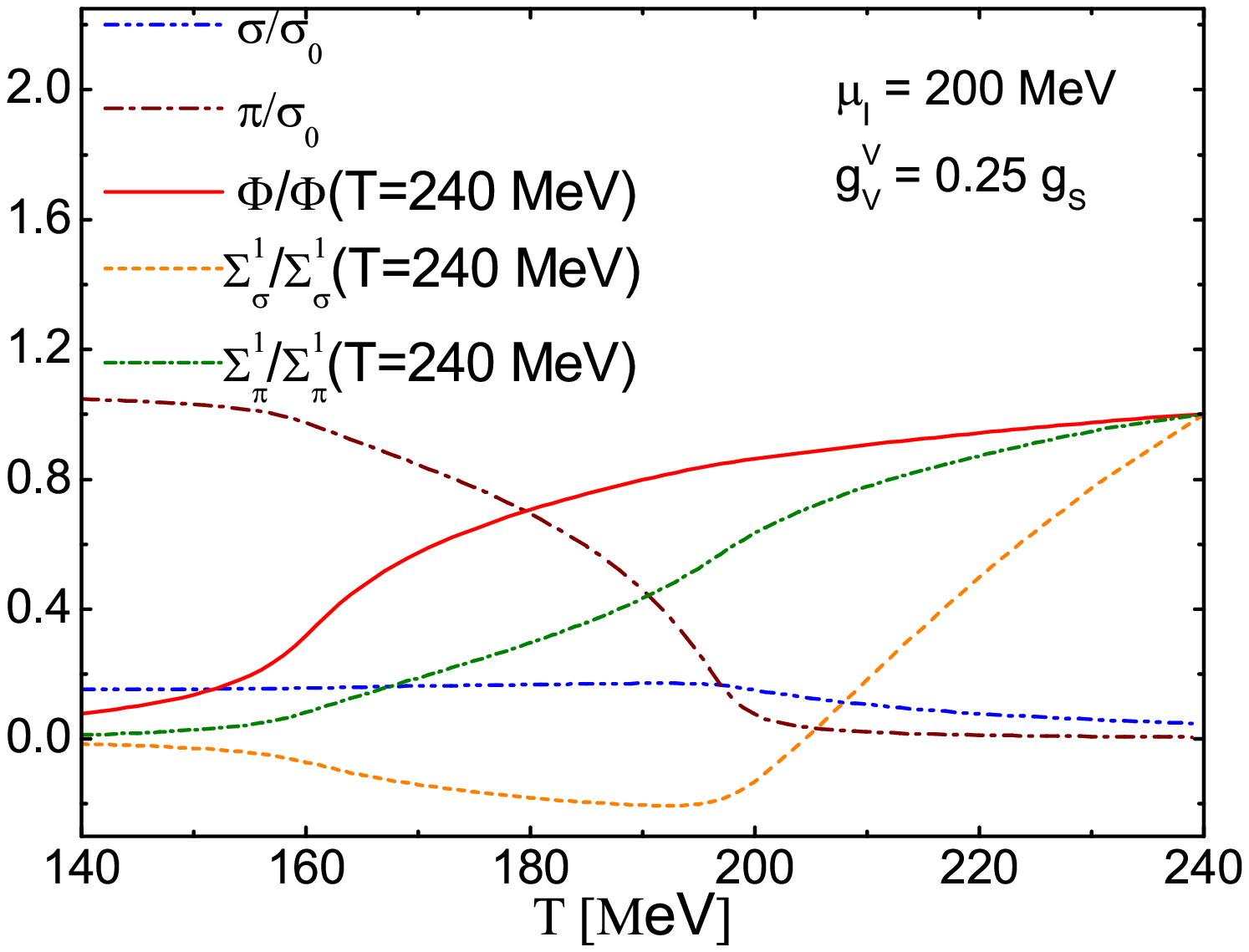} \vspace{1em}\\
\end{center}
\caption{ The temperature dependences of the normalized conventional ployakov loop,
quark and pion condensates and their corresponding dual parters at $\mu_I=200 \MeV$.
 }
\label{fig:plcondensate2}
\end{figure}

\begin{figure}[t]
\begin{center}
\includegraphics[width=1.0\columnwidth]{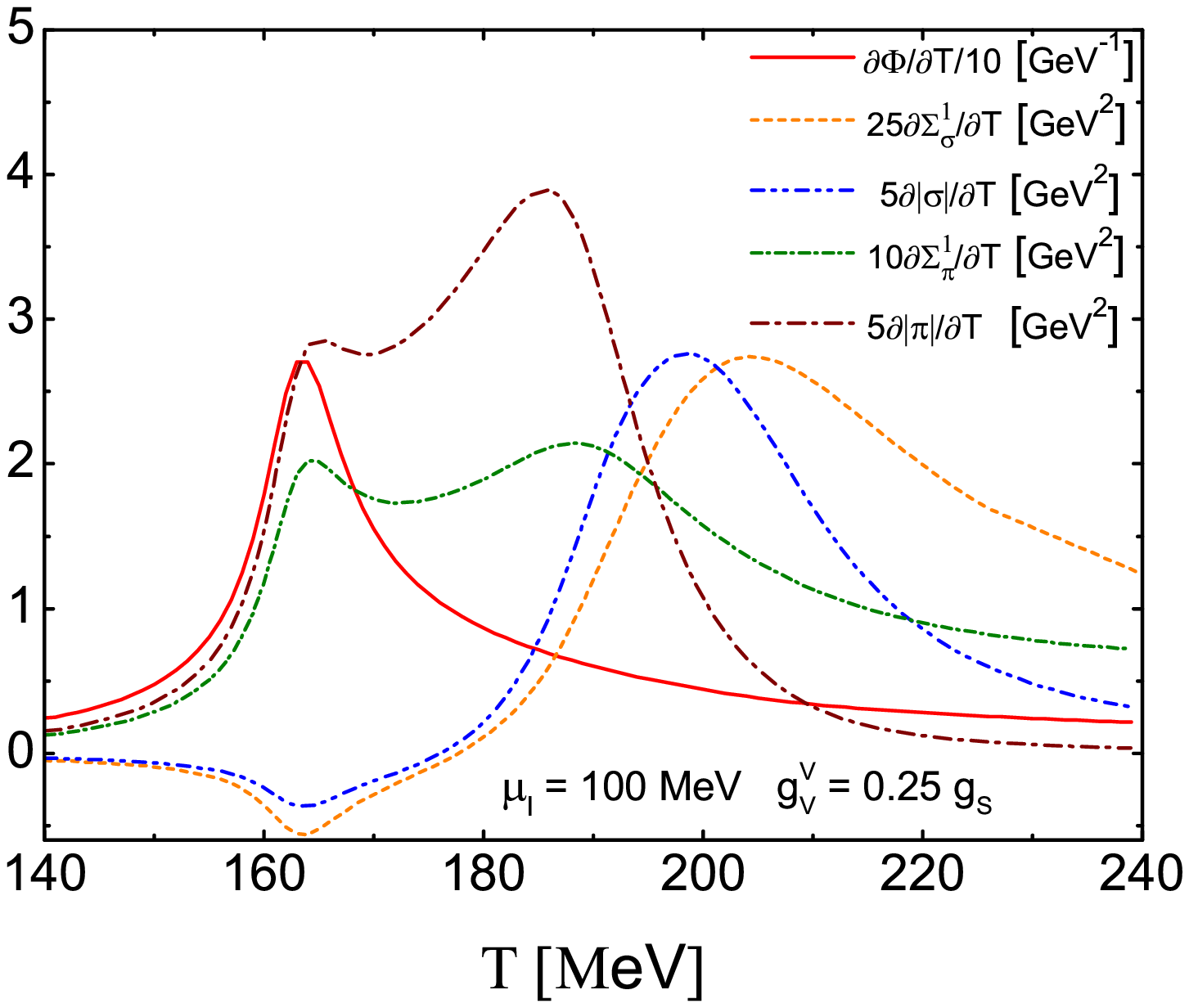} \vspace{1em}\\
\end{center}
\caption{ The temperature dependences of the T-derivatives of the conventional ployakov loop,  quark and
pion condensates and their corresponding dual parters at $\mu_I=100 \MeV$.
}
\label{fig:sus1}
\end{figure}

\begin{figure}[t]
\begin{center}
\includegraphics[width=1.0\columnwidth]{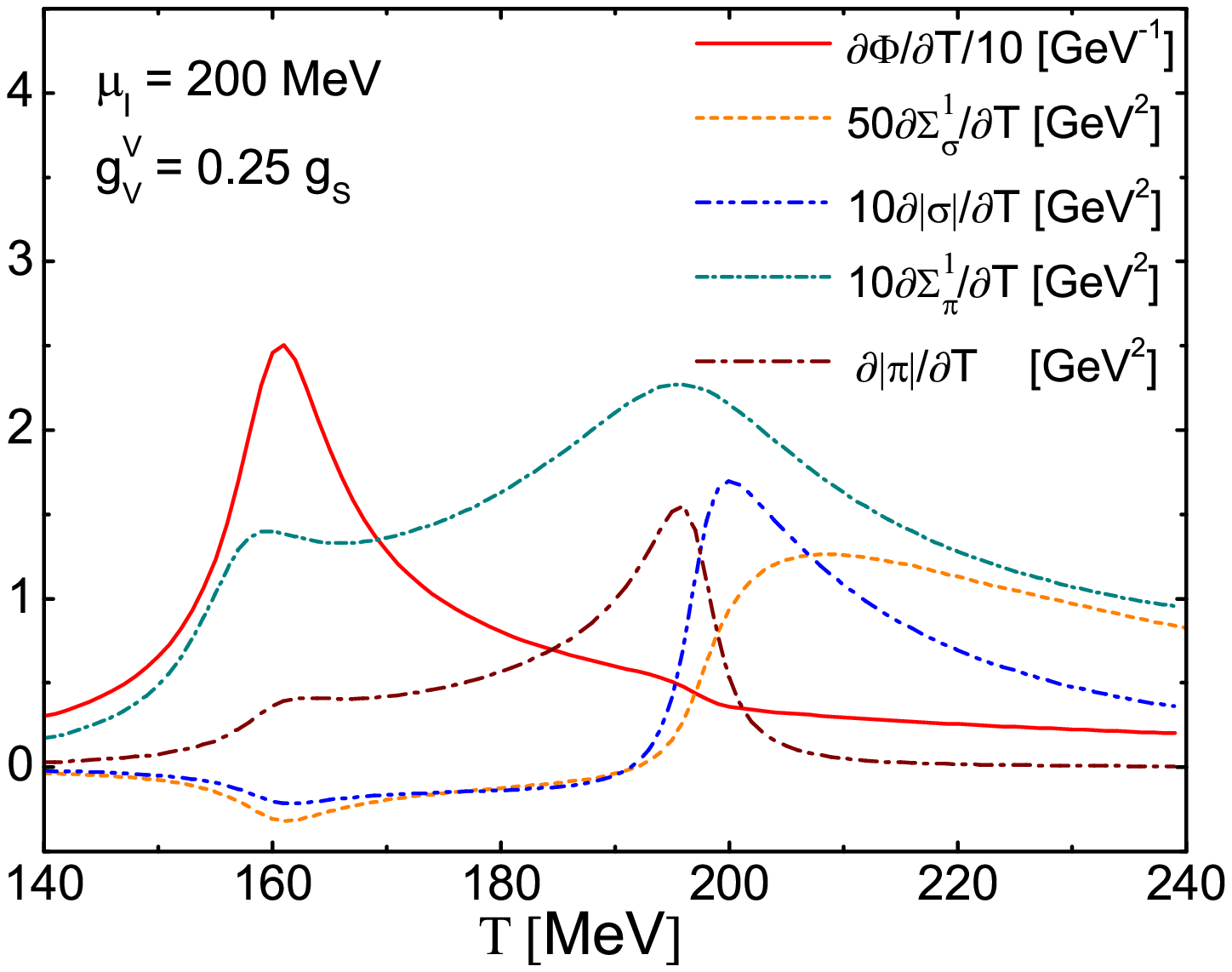} \vspace{1em}\\
\end{center}
\caption{ The temperature dependences of the T-derivatives of the conventional ployakov loop,  quark and
pion condensates and their corresponding dual parters at $\mu_I=200 \MeV$.
}
\label{fig:sus2}
\end{figure}

\subsection{ Model parameters }

In our calculation, the model parameters related to NJL, such as the current quark mass $m$, the momentum cutoff
$\Lambda$ and the scalar coupling $g_s$ are all adopted from \cite{Zhang:2006gu}, which take the values
\begin{equation}
m=5.5\mathrm{MeV},\Lambda=0.651\mathrm{GeV}, \text{and}\quad g_s=5.04 \mathrm{GeV}^{-2},
\end{equation}
respectively. The vector coupling $g_v^v$ is fixed as $0.25g_s$, which is supported by the instanton
liquid molecular model.

As for the PL potential, we employ the logarithm form \cite{Roessner:2006xn}. It has been reported that
this type of $\mathcal{U}$ can reproduce the LQCD data at finite imaginary chemical potential, but the
polynomial one (which is used in \cite{Zhang:2006gu}) does not \cite{Kashiwa:2009ki}. Following
\cite{Kashiwa:2009ki}, the parameter $T_0$ in the logarithm potential is fitted as $200 \MeV$ to
reproduce the lattice pseudo-critical temperature $T_c$ at zero density.

For computational convenience, a small $\lambda$ with value $m/100$ is used. It is confirmed that the
deviation of our main results from zero $\lambda$ is negligible.

\section{ Numerical results and discussions }

\subsection{ $\phi$-dependence of quark and pion condensates }

The generalized quark and pion condensates as functions of $\phi$ for $\mu_I=100\,\text{MeV}$ at different 
temperatures are shown in Fig.~\ref{fig:dualc1}. Fig.~\ref{fig:dualc1}.a indicates that the shapes of
$|\sigma(\phi)|$ for temperatures below and above $T_c^{\chi}$ are quite different:
for $T=210\,\text{MeV}$ and $240\, \text{MeV}$ (or $T>T_c^{\chi}$), the quark condensates are concave lines with 
$|\sigma(\pi)|<|\sigma(0)|$; but for $T=150\,\text{MeV}$ and $170\,\text{MeV}$ (or $T<T_c^{\chi}$), they are convex
ones with $|\sigma(\pi)|>|\sigma(0)|$. The transition line between the concave and convex ones takes the wavy 
shape, as displayed in Fig.~\ref{fig:dualc1}.a for $T=190\,\text{MeV}$ (or $T{\sim}T_c^{\chi}$). All this is quite 
different from what obtained in \cite{Bilgici:2008qy,Fischer:2009gk,Kashiwa:2009ki}, where only concave curves emerge 
for vanishing $\mu$ and $\mu_I$. Fig.~\ref{fig:dualc1}.a also shows that $|\sigma(\phi\sim\pi)|$ first increases and 
then decreases with $T$ due to the impact of pion condensate
\footnote{This is also observed in \cite{Zhang:2006gu} and other chiral model studies. The reason for such
an anomaly is that the quantity $\sqrt{\sigma^2+\pi^2}$ always decreases with $T$ but $|\pi|$ reduces more
quickly since $\lambda$ is zero but $m$ is finite.}; but in \cite{Bilgici:2008qy,Fischer:2009gk,Kashiwa:2009ki},
$|\sigma(\phi\sim\pi)|$ always decreases with $T$.

In contrast, Fig.~\ref{fig:dualc1}.b shows that all lines of $|\pi(\phi)|$ at different fixed $T$ are concave
curves. We see that in the fermionic-like region (namely the area for $\phi\sim\pi$), $|\pi(\phi)|$ decreases with 
$T$ but it increases with $T$ in the bosonic-like region (namely the area for $\phi$ near zero or $2\pi$). In 
addition, the curve of $|\pi(\phi)|$ becomes more flat with decreasing $T$. All those is very similar to the 
$\phi$-dependence of the quark condensate obtained for zero $\mu$ and $\mu_I$
\cite{Bilgici:2008qy,Fischer:2009gk,Kashiwa:2009ki}. The similarity can be understood in the following way:
For $\mu_I>m_\pi/2$, the quark condensate partially turns into the pion condensate, and thus the later inherits 
some properties of the former. However, the $\phi$-dependence of $|\sigma(\phi)|$ changes obviously due to such 
a transformation, as shown in Fig.~\ref{fig:dualc1}.a.

Figure.~\ref{fig:dualc2} shows the $\phi$-dependence of quark and pion condensates at different temperatures for
$\mu_I=200\,\text{MeV}$. Compared with Fig.~\ref{fig:dualc1}, we see that $\sigma(\phi)$ is suppressed and $\pi(\phi)$ 
is enhanced (suppressed) in the fermion-like (boson-like) region. But the $\phi$ and $T$ dependences of these two 
condensates  are still qualitatively consistent with that displayed in Fig.~\ref{fig:dualc1}.

\subsection{ Thermal behaviors of dual condensates }

Two dual condensates and other three (pseudo-) order parameters as functions of $T$ for $\mu_I=100\,\text{MeV}$
and $200\,\text{MeV}$ are shown in Figs.~\ref{fig:plcondensate1}-\ref{fig:plcondensate2}, respectively. The quark 
and pion condensates are obtained with physical boundary condition
$\phi=\pi$, which are normalized by $\sigma_0=\sigma(T=0,\mu=0,\mu_I=0)$; the dual condensates and PL are normalized by
their corresponding values at $T=240\,\text{MeV}$. We mainly focus on the thermal behaviors of these quantities near the
phase transitions.

Figure.~\ref{fig:plcondensate1} shows that $|\pi(\phi=\pi)|$ ($\Phi$) decreases (increases) monotonically with $T$, but
$|\sigma(\phi=\pi)|$ first increases (slowly) up to $T\sim180\,\text{MeV}$ and then decreases. The similar $T$-dependences
of these quantities are also observed in Fig.~\ref{fig:plcondensate2}. These results are qualitatively agreement with what obtained
in \cite{Zhang:2006gu} by using the polynomial PL potential. As mentioned above, the increase of $|\sigma(\phi=\pi)|$ with
$T$ is due to the competition between the quark and pion condensates.

Consistent with Fig.~\ref{fig:dualc1}.b, Fig.~\ref{fig:plcondensate1} indicates that the normalized DPC really behaves like an
order parameter for center symmetry: analogous to the DPL obtained in \cite{Bilgici:2008qy,Fischer:2009gk,Kashiwa:2009ki}, it keeps 
rather small value in low temperature region and gradually becomes larger with $T$. Like the thin PL, the DPC increases monotonically
with $T$. However, the normalized DPL in Fig.~\ref{fig:plcondensate1} shows abnormal thermal behavior, which first reduces with $T$
(up to $T\sim180 \text{MeV}$) and then raises. Fig.~\ref{fig:plcondensate1} also shows that the DPL even becomes negative near
and below $T\sim190 \text{MeV}$. The similar $T$-dependences of DPL and DPC are also observed in Fig.~\ref{fig:plcondensate2}.

The abnormal thermal behavior of DPL can be traced back to the non-concave lines of $\sigma(\phi)$ displayed in
Figs.\ref{fig:dualc1}(a)-\ref{fig:dualc2}(a). As mentioned, due to the influence of $\pi(\phi)$, $|\sigma(\phi)|$ increases with 
$T$ below $T_c^{\pi}$ (the critical temperature for the $I_3$ symmetry restoration) for $\phi\sim\pi$. This results in the DPL not 
always raising with $T$. Actually, Figs.~\ref{fig:plcondensate1}-\ref{fig:plcondensate2} clearly show that when $|\sigma(\phi=\pi)|$ 
increases with $T$, the DPL decreases, and vice versa. So the DPL is quite sensitive to the $T$-dependence of the quark 
condensate. In contrast, Figs.~\ref{fig:plcondensate1}-\ref{fig:plcondensate2} indicate that the DPL is insensitive to the thin 
PL, at least at low temperature region. All these suggests that the DPL obtained with physical quark masses mainly reflects 
chiral transition rather than deconfinement. Such a conclusion is agreement with the claim given in \cite{Benic:2013zaa} that 
the rapid change of DPL near $T_c^{\chi}$ in NJL is totally driven by the chiral restoration.

Following Ref.~\cite{Benic:2013zaa}, we also calculate several susceptibilities which are defined as the $T$ derivatives 
of the quantities displayed in Figs.~\ref{fig:plcondensate1}-\ref{fig:plcondensate2}. As in \cite{Benic:2013zaa}, 
the peak of a susceptibility is used to locate the critical temperature. The susceptibilities as functions of $T$ for 
$\mu_I=100 \MeV$ and $200 \MeV$ are shown in Fig.~\ref{fig:sus1} and Fig.~\ref{fig:sus2}, respectively. 

Fig.~\ref{fig:sus1} shows that the PL susceptibility has only one peak, which indicates $T_c^{P}=163\,\text{MeV}$. 
But the other susceptibilities all have double peaks, one of which coincides with $T_c^{P}$ due to the coupling between 
the PL and quark/pion condensate in PNJL. We see that the highest peaks of $\partial{\Sigma_\sigma}^{1}/\partial{T}$ 
and $\partial{|\sigma|}/\partial{T}$  in Fig.~\ref{fig:sus1} are very close to each other, and the corresponding critical 
temperatures $T_c^{d\sigma}$ and $T_c^{\sigma}$ are about $40\,\text{MeV}$ lager than $T_c^{P}$. The coincidence of 
$T_c^{d\sigma}$ and $T_c^{\sigma}$ is consistent with \cite{Benic:2013zaa} even the PL dynamics and pion condensate are 
considered in our calculations. In addition, Fig.~\ref{fig:sus1} also shows that the critical temperatures $T_c^{d\pi}$ and $T_c^{\pi}$
(extracting from $\partial{\Sigma_\pi}^{1}/\partial{T}$ and $\partial{|\pi|}/\partial{T}$, respectively) almost coincide, which
are about $25 \MeV$ lager than $T_c^{P}$. The slight difference between $T_c^{\pi}$ and $T_c^{d\pi}$ can be traced back
to a small $\lambda=0.1m$ used in our numerical calculation. All these coincidences are also observed in Fig.~\ref{fig:sus2}. 
Actually, the accordance of $T_c^{\pi}$ and $T_c^{d\pi}$ shown in Figs.~\ref{fig:plcondensate1}-\ref{fig:plcondensate2} 
is completely analogous to $T_c^{\sigma}=T_c^{d\sigma}$ obtained in \cite{Benic:2013zaa} for zero m.  We thus get the 
similar conclusion that the rapid change of DPC near $T_c^{\pi}$ is driven by the restoration of $I_3$ symmetry, even 
it exhibits an order parameter-like thermal behavior as the PL.

Note that we also perform the similar calculations by varying $g_v^v$ in this model. We confirm that the thermal properties
of DPL and DPC shown in Figs.1-6 do not change qualitatively.

\subsection{ The case in the chiral limit }

Here we only show the results with physical quark masses. In the chiral limit with finite $\mu_I$, the pion condensate 
appears at low temperatures but the quark condensate vanishes. Or in other words, the quark condensate totally turns into 
the pion condensate due to the nonzero $\mu_I$. Correspondingly, the dual quark condensate is replaced by the dual pion condensate. 
In this case, the approach used in \cite{Benic:2013zaa} for analysing the DPL, such as the Ginzburg-Landau method, can be borrowed 
directly to study the DPC. We have checked that without the quark condensate, the $T$-dependence of DPC for finite $\mu_I$ is 
much more similar to that of DPL obtained in the chiral limit at zero $\mu_I$ \cite{Benic:2013zaa}. Actually, this case is 
also analogous to the situation with physical quark masses and large $\mu_I$, in which the pion condensate dominates and the 
suppressed quark condensate can be ignored.  

\subsection{ Discussions }

So beyond \cite{Benic:2013zaa}, we give further evidences that the dual observable may not really reflect deconfinement transition, 
even it is constructed from center symmetry. Note that in \cite{Benic:2013zaa}, the author still insists that the DPL calculated 
in other methods, such as the formalism of truncated Dyson-Schwinger Equation \cite{Fischer:2011mz}, can be used as an order 
parameter for deconfinement. However, the dual quark condensates obtained in NJL \cite{Benic:2013zaa} and DSE \cite{Fischer:2011mz}
are qualitatively consistent with each other. In addition, our result and Ref.\cite{Kashiwa:2009ki} all suggest that the $T$-dependence 
of DPL obtained in PNJL is also quite similar to that calculated in the conventional NJL. So if the rapid change of DPL with $T$ 
merely indicates the chiral transition in NJL, there is no reason to explain it as the deconfinement transition in other methods. 
We thus argue that the so called coincidence of chiral and deconfinement transitions in other formalisms using the DPL as the 
order parameter may also be an antifact, just as that in NJL \cite{Mukherjee:2010cp}.

Here we stress that the statement that the DPL only reflects the chiral transition given in \cite{Benic:2013zaa} may hold 
true not only in NJL but also in QCD with physical quark masses
\footnote{Actually, a general method, the Ginzburg-Landau analysis is used in \cite{Benic:2013zaa}
to prove the coincidence of the chiral transition and the rapid change of DPL.}. We argue that the main reason responsible
for it should be the severe violation of center symmetry due to the light dynamical quarks. We point out that it is a natural
result that the dual observables, such as the DPL, contain very limited information on confinement-deconfinement
transition, unless the dynamical quarks are heavy enough. Actually, the abnormal $T$-dependence of DPL displayed in
Figs.~\ref{fig:plcondensate1}-\ref{fig:plcondensate2} just reflects that this quantity is a bad order parameter
for deconfinement with physical quark masses.

In the heavy quark limit, the DPL approaches the PL \cite{Bilgici:2008qy}, and both are good order parameters for center symmetry.
But in the light quark limit, the DPL differs significantly from the PL. So even the PL has been widely adopted to serve as the order 
parameter for deconfinement transition, it does not mean that the DPL and other dual observables can do this too with physical quark masses.
Actually, even to what extent does the PL contain the information of quark deconfinement transition is also a subtle problem if the quark
mass is very small. Formally, the definitions of the DPL and DPC involve quark fields which are naturally related to the quark and pion
condensates, respectively. So it is not strange that the rapid change of the former exactly reflects the chiral transition in the chiral
limit and that of the later just indicates the restoration of $I_3$ symmetry. Thus one should be cautious to mention the so called 
coincidence of the chiral and deconfinement transitions through the study of dual observables or other quantities constructed from 
center symmetry.

\section{ Conclusion }

The dual observables as possible order parameters for center symmetry are tested at finite temperature and isospin density with
physical quark masses. Besides the dressed Polyakov-loop, another simple dual condensate, namely the dual pion condensate is
proposed for $\mu_I>m_\pi/2$. We investigate the thermal behaviors of these two quantities in the PL enhanced NJL model of QCD
by considering the pion condensation. Our conclusion is that both dual observables contain little information on quark
deconfinement transition.

First, we find that the twisted angle dependence of pion condensate is quite analogous to that of quark condensate obtained at
zero $\mu$ and $\mu_I$. Correspondingly, the DPC exhibits the similar $T$-dependence as the conventional PL for $\mu_I>m_\pi/2$.
We demonstrate that the derivative of DPC with respect to $T$ peaks exactly at $T_c^\pi$ at which the pion condensation evaporates. 
This is very similar to the coincidence verified in \cite{Benic:2013zaa}, where the critical temperature extracting from the 
DPL equals exactly to $T_c^\chi$ in the chiral limit. Thus, we get the analogous conclusion that the rapid change of DPC near 
$T_c^\pi$ is driven by the restoration of $I_3$ symmetry. So even the DPC shows order parameter-like behavior, the critical 
temperature extracted from it has nothing to do with the deconfinement transition.

Second, we find that the DPL displays abnormal thermal property for $\mu_I>m_\pi/2$, which even decreases with $T$ for
$T\leq T_c^\chi$. This is quite different from the thin PL, which always increases with $T$. The anomaly arises due to the interplay 
between the quark and pion condensates. We verify that the DPL increases with $T$ if the quark condensate (its absolute value) 
decreases, and vice versa. This implies the variation of DPL with $T$ is mainly determined by the chiral dynamics rather than 
the confinement in the situation with physical quark masses, which is in agreement with \cite{Benic:2013zaa}.

We thus conclude that both dual condensates are unqualified order parameters for deconfinement transition, even they are constructed
from center symmetry. Different from \cite{Benic:2013zaa}, we argue that this conclusion holds not only in the NJL-type 
model, but also in QCD because of the severe violation of center symmetry. We stress that if the quark mass is very small, the dual observable 
should naturally loses its role as an effective indicator for deconfinement. We thus suspect that the so called coincidence of 
the chiral and deconfinement phase transitions obtained from the thermal properties of dual observables may be just an artifact.

Note that recently the temporally odd lattice has been used to investigate the relation between chiral symmetry breaking and confinement
\cite{Doi:2014zea,Iritani:2013pga}. It is found that the low-lying Dirac zero mode has little contribution to the PL, which indicates  
there is no one-to-one correspondence between the chiral symmetry breaking and confinement in QCD. This conclusion disfavors the prior 
lattice results that the chiral restoration and deconfinement happen at the same temperature at zero density. So whether the chiral and 
deconfinement transitions coincides or not is still under debate and need further study.

Since there is no sign problem at finite $\mu_I$, our study can be performed in the lattice calculation. In addition, the pion condensation
has been investigated in the DSE formalism \cite{Zhang:2006dn} and other effective models of QCD, such as the quark-meson model. It is also 
interesting to investigate the thermal properties of dual condensates at finite $\mu_I$ within these methods.       

\vspace{5pt}
\noindent{\textbf{\large{Acknowledgements}}}\vspace{5pt}\\
Z.Z. was supported by the NSFC ( No.11275069 ).


\begin{thebibliography}{}



\bibitem{Aoki:2009sc}
  Y.~Aoki, S.~Borsanyi, S.~Durr, Z.~Fodor, S.~D.~Katz, S.~Krieg and K.~K.~Szabo,
  JHEP {\bf 0906}, 088 (2009).


\bibitem{Borsanyi:2010bp}
  S.~Borsanyi {\it et al.}  [Wuppertal-Budapest Collaboration],
  JHEP {\bf 1009}, 073 (2010).

\bibitem{Banks:1980}
  T.~Banks and A.~Casher,
  Nucl.\ Phys.\ B{\bf 169}(1980)103

\bibitem{Gattringer:2006ci}
  C.~Gattringer,
  Phys.\ Rev.\ Lett.\  {\bf 97}, 032003 (2006).


\bibitem{Bilgici:2008qy}
  E.~Bilgici, F.~Bruckmann, C.~Gattringer and C.~Hagen,
  Phys.\ Rev.\ D {\bf 77}, 094007 (2008).

\bibitem{Bilgici:2009tx}
  E.~Bilgici, F.~Bruckmann, J.~Danzer, C.~Gattringer, C.~Hagen, E.~M.~Ilgenfritz and A.~Maas,
  Few Body Syst.\  {\bf 47}, 125 (2010).
  
\bibitem{Zhang:2010ui}
  B.~Zhang, F.~Bruckmann, C.~Gattringer, Z.~Fodor and K.~K.~Szabo,
  AIP Conf.\ Proc.\  {\bf 1343}, 170 (2011)
  [arXiv:1012.2314 [hep-lat]].
  

\bibitem{Synatschke:2007yt}
  F.~Synatschke, A.~Wipf and C.~Wozar,
  Phys.\ Rev.\ D {\bf 75}, 114003 (2007).


\bibitem{Synatschke:2008yt}
  F.~Synatschke, A.~Wipf and K.~Langfeld,
  Phys.\ Rev.\ D {\bf 77}, 114018 (2008).

\bibitem{Braun:2009gm}
  J.~Braun, L.~M.~Haas, F.~Marhauser and J.~M.~Pawlowski,
  Phys.\ Rev.\ Lett.\  {\bf 106}, 022002 (2011).

\bibitem{Fischer:2009wc}
  C.~S.~Fischer,
  Phys.\ Rev.\ Lett.\  {\bf 103}, 052003 (2009).

\bibitem{Fischer:2009gk}
  C.~S.~Fischer and J.~A.~Mueller,
  Phys.\ Rev.\ D {\bf 80}, 074029 (2009).

\bibitem{Fischer:2011mz}
  C.~S.~Fischer, J.~Luecker and J.~A.~Mueller,
  Phys.\ Lett.\ B {\bf 702}, 438 (2011).

\bibitem{Fischer:2010fx}
  C.~S.~Fischer, A.~Maas and J.~A.~Muller,
  Eur.\ Phys.\ J.\ C {\bf 68}, 165 (2010).


\bibitem{Kashiwa:2009ki}
  K.~Kashiwa, H.~Kouno and M.~Yahiro,
  Phys.\ Rev.\ D {\bf 80}, 117901 (2009).


\bibitem{Gatto:2010qs}
  R.~Gatto and M.~Ruggieri,
  Phys.\ Rev.\ D {\bf 82}, 054027 (2010).

\bibitem{Mukherjee:2010cp}
  T.~K.~Mukherjee, H.~Chen and M.~Huang,
  Phys.\ Rev.\ D {\bf 82}, 034015 (2010).

\bibitem{Benic:2013zaa}
  S.~Benic,
  Phys.\ Rev.\ D {\bf 88}, 077501 (2013).


\bibitem{Son:2001}D. T. Son and M. A. Stephanov. Phys. Rev. Lett. {\textbf{86 }}, 592 (2001);
Phys. At. Nucl. {\textbf{64}}, 834 (2001).

\bibitem{Kogut:2002}J. B. Kogut and D. K. Sinclair, Phys. Rev. D {\textbf{66}}, 034505 (2002);

\bibitem{Kogut:2004}J. B. Kogut and D. K. Sinclair,Phys. Rev. D {\textbf{70 }}, 094501 (2004).


\bibitem{Zhang:2006gu}
  Z.~Zhang and Y.~-X.~Liu,
  Phys.\ Rev.\ C {\bf 75}, 064910 (2007).
\bibitem{Zhang:2013oia}
  Z.~Zhang and H.~P.~Su,
  Phys.\ Rev.\ D {\bf 89}, 054020 (2014)
\bibitem{Roessner:2006xn}
  S.~Roessner, C.~Ratti and W.~Weise,
  Phys.\ Rev.\ D {\bf 75}, 034007 (2007)


\bibitem{Doi:2014zea}
  T.~M.~Doi, H.~Suganuma and T.~Iritani,
  Phys.\ Rev.\ D {\bf 90}, no. 9, 094505 (2014)
  [arXiv:1405.1289 [hep-lat]].

\bibitem{Iritani:2013pga}
  T.~Iritani and H.~Suganuma,
  PTEP {\bf 2014}, no. 3, 033B03 (2014).

\bibitem{Zhang:2006dn}
  Z.~Zhang and Y.~x.~Liu,
  Phys.\ Rev.\ C {\bf 75}, 035201 (2007).
  
 


\end{thebibliography}
\end{document}